# Comparative Study of Chemically Synthesized and Exfoliated Multilayer MoS$_2$ Field-Effect Transistors

Wan Sik Hwang[1,3,a)], Maja Remskar[2], Rusen Yan[3], Tom Kosel[3], Jong Kyung Park[4], Byung Jin Cho[4], Wilfried Haensch[1], Huili (Grace) Xing[3], Alan Seabaugh[3], and Debdeep Jena[3,b)]

[1]IBM T. J. Watson Research Center, Yorktown Heights, NY 10598, USA

[2]Solid State Physics department, Jozef Stefan Institute, Jamova 39, SI-1000 Ljubljana, Slovenia

[3]Department of Electrical Engineering, University of Notre Dame, Notre Dame, IN 46556, USA

[4] Department of Electrical Engineering, Korea Advanced Institute of Science and Technology, Daejeon, 305-701, Korea

- a) whwang@us.ibm.com or whwang1@nd.edu,
- b) djena@nd.edu

**ABSTRACT**

We report the realization of field-effect transistors (FETs) made with chemically synthesized multilayer 2D crystal semiconductor MoS$_2$. Electrical properties such as the FET mobility, subthreshold swing, on/off ratio, and contact resistance of chemically synthesized (s-) MoS$_2$ are indistinguishable from that of mechanically exfoliated (x-) MoS$_2$, however flat-band voltages are different, possibly due to polar chemical residues originating in the transfer process. Electron diffraction studies and Raman spectroscopy show the structural similarity of s-MoS$_2$ to x-MoS$_2$. This initial report on the behavior and properties of s-MoS$_2$ illustrates the feasibility of electronic devices using synthetic layered 2D crystal semiconductors.

Two-dimensional (2D) crystal materials are receiving increased attention for future electronic devices. Short-channel effects in modern transistors originate from the 3-dimensional (3D) nature of the gate control; introduction of 2D materials in the channel significantly improves the gate electrostatics. Graphene is a true 2D material [1], but its lack of a bandgap results in high leakage in the off-state in conventional transistor geometries. Alternative switching mechanisms [2 ~ 4] could enable electronic switching with high on-off current ratios. 2D transition-metal dichalcogenide (TMD) materials such as $MoS_2$ [5, 6], $WSe_2$ [7, 8] and $WS_2$ [9, 10] have drawn considerable attention due to the presence of a bandgap, in contrast to graphene. Prior studies of TMD multilayered materials have been conducted using exfoliated layers [5~9], similar to the initial work with graphene [1]. The size and quality of naturally occurring exfoliated layered semiconductor materials are limited and uncontrollable, so it is important to develop synthetic techniques. Multilayer TMDs, like graphite, are excellent solid lubricating materials and have been chemically synthesized in large volumes, but have not been investigated intensively for transistors [11 ~ 13] except in very recent work [14]. Here we report the fabrication and demonstration of chemically synthesized multilayer (s-) $MoS_2$ FETs and compare the properties with exfoliated (x-) $MoS_2$ FETs.

s-$MoS_2$ flakes were grown by an iodine-transport method from previously synthesized $MoS_2$ (0.6 g) at 1060 K in an evacuated silica ampoule at a pressure of $10^{-3}$ Pa, and with temperature gradient of 6.8 K/cm. The volume concentration of iodine was 11 mg/cm$^3$. After 21 days of growth, the silica ampoule was slowly cooled to room temperature at a rate of 30 °C / hour. The $MoS_2$ flakes were then dispersed by sonication in isopropyl alcohol (IPA) and transferred onto a 30 nm thick atomic-layer-deposited (ALD) $Al_2O_3$ dielectric on a $p$+Si substrate held at 100 °C until dry. For comparison, x-$MoS_2$ flakes were released and transferred from bulk MoS$_2$ using scotch tape. The height of the MoS$_2$ flakes are in the range of 20 ~ 40 nm. Source and drain contacts were defined by electron beam lithography (EBL) using Ti/Au (5/100 nm) contacts. The devices were annealed at 300 °C for 3 hours under Ar/He flow to decrease the contact resistance. A schematic cross-sectional image of the back-gated (BG) MoS$_2$ device is shown in the inset of Fig. 1(a).

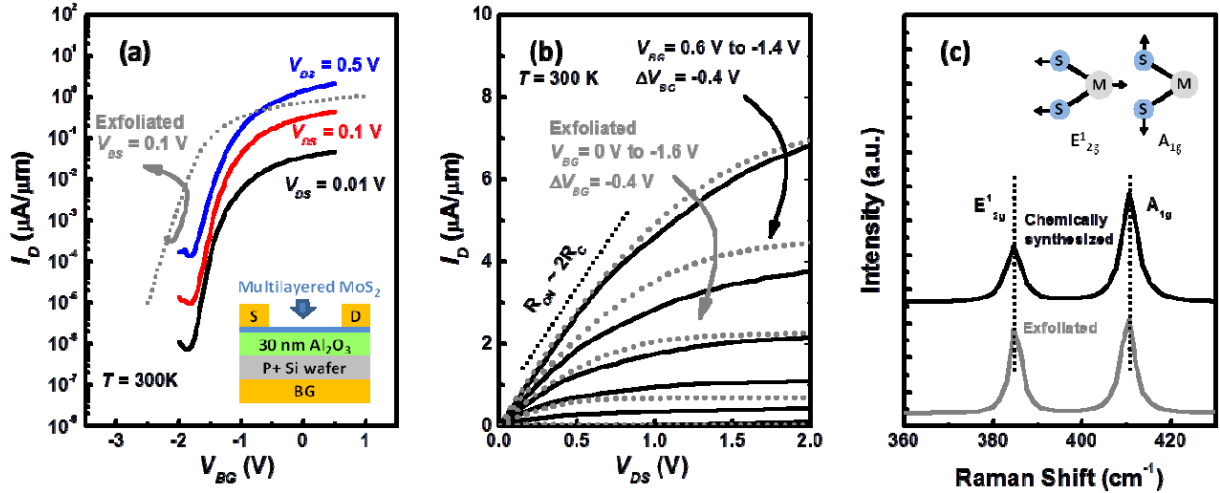

**FIG. 1**. Transport properties and Raman spectroscopy of FETs of s-MoS$_2$ with W/L = 1/2 μm and x-MoS$_2$ with W/L = 5/1.5 μm. (a) Drain current, $I_D$, per unit gate width vs. back-gate voltage, $V_{BG}$, of s-MoS$_2$ at various drain voltages, $V_{DS}$. The transfer curve of an x-MoS$_2$ is also shown for comparison, dotted line. (b) Common source transistor characteristics comparing s-MoS$_2$ (solid lines) and x-MoS$_2$ (dashed lines) FETs. (c) Raman spectra ($\lambda_{exc}$ = 488 nm) of both s-MoS$_2$ and x-MoS$_2$ materials with a laser power of 1.5 mW and a spot size of 0.5-1 μm. The inset sketch shows the two primary vibrational modes in MoS$_2$ leading to the two peaks in the Raman spectrum.

Figure 1(a) shows the measured drain current, $I_D$, per unit gate width, versus the back-gate-to-source voltage, $V_{BG}$, at room temperature for a multilayer s-MoS$_2$ channel at three drain biases. The gate modulation is ~$10^5$ and the gate leakage current is much lower (less than 1 pA/μm) than the drain current. This large gate modulation relative to graphene is attributed to the presence of a bandgap. The device shows clear *n*-type behavior indicating accumulation of electrons (*n*-type conductivity) for positive back-gate bias. The comparative transfer curves of x-MoS$_2$ FETs are also shown in Fig 1(a). s-MoS$_2$ and x-MoS$_2$ FETs show highly similar transfer characteristics. The extracted field-effect mobilities of both x-MoS$_2$ and s-MoS$_2$ FETs are ~30 cm$^2$/Vs at room temperature. The subthreshold swing (SS) of the x-MoS$_2$ FET is 200 mV/dec. and that of the s-MoS$_2$ FET is 190 mV/dec. The average values of the FET mobility of each type of both x-MoS$_2$ and s-MoS$_2$ are ~15 cm$^2$/Vs and the SS of those are ~170 mV/dec. Another s-MoS$_2$ (W/L = 1/2 μm) FET is compared with x-MoS2 and electrical properties still work out to be similar.

The subthreshold swing is similar for the s-MoS$_2$ and x-MoS$_2$ FETs, but higher than the ideal Boltzmann limit of 60 mV/decade. The similar SS suggests that the interface charge leading to the higher subthreshold swing likely arises from traps in the Al$_2$O$_3$ dielectric, which are identical for the s-MoS$_2$ and x-MoS$_2$ FETs. The major difference between s-MoS$_2$ and x-MoS$_2$ FETs is in the threshold voltage. The flat-band voltage of the x-MoS$_2$ FETs is higher than that of s-MoS$_2$ by approximately 1 V. The possible reasons for this shift could be a) different unintentional doping densities in s-MoS$_2$ and x-MoS$_2$, or b) scotch tape residue-induced-charges possibly leading to a higher flat-band voltage for the x-MoS$_2$ compared to the s-MoS$_2$ FETs, which do not experience the tape exfoliation procedure. Identifying the precise reason for this threshold shift requires further work.

The family of $I_D$ - $V_{DS}$ curves at various $V_{GS}$ in Fig 1(b) shows typical transistor behavior including a linear increase of current at low $V_{DS}$ and current saturation at high $V_{DS}$. The behavior shows desirable transistors attributes such as ohmic contacts, current saturation, and good gate electrostatic control. However, the current levels are in the µA/µm regime for micron long gate-lengths, which is low. The reason for the low current is a high contact resistance. The contact resistances of both s-MoS$_2$ and x-MoS$_2$ FETs in this work were extracted to be ~80 Ω mm at low $V_{DS}$. This value is comparable to the ~69 reported for exfoliated MoS$_2$ FETs [6]. This is an extremely high value, and currently holds back the performance of the FETs. The observation here is that s-MoS$_2$ and x-MoS$_2$ FETs have similar contact resistances, and are both high. A significant increase in the current drive of the FETs is expected if the contact resistance can be lowered to 1 Ω mm regime or lower, as is the chase in Si and III-V FETs. The contact resistance can be lowered using a low work function metal like Sc [15].

The measured Raman spectra shown in Fig. 1(c) at an excitation wavelength, $\lambda_{exc}$, of 488 nm is highly similar between s-MoS$_2$ and x-MoS$_2$ flakes. The spectrum exhibits two peaks: one in the $E_{2g}^1$ range, corresponding to in-plane vibrations at ~385 cm$^{-1}$, and the other in the $A_{1g}$ range corresponding to out-of-plane vibrations at ~410 cm$^{-1}$. Since the Raman spectrum and transport properties are almost identical for s-MoS$_2$ and x-MoS$_2$, this confirms the overall similarity of chemically-synthesized and exfoliated MoS$_2$. In order to further investigate the atomic properties of s-MoS$_2$, transmission electron microscopy (TEM) was performed. The information obtained from s-MoS$_2$ is shown in Fig. 2 and compared with that of x-MoS$_2$ in Fig. 3.

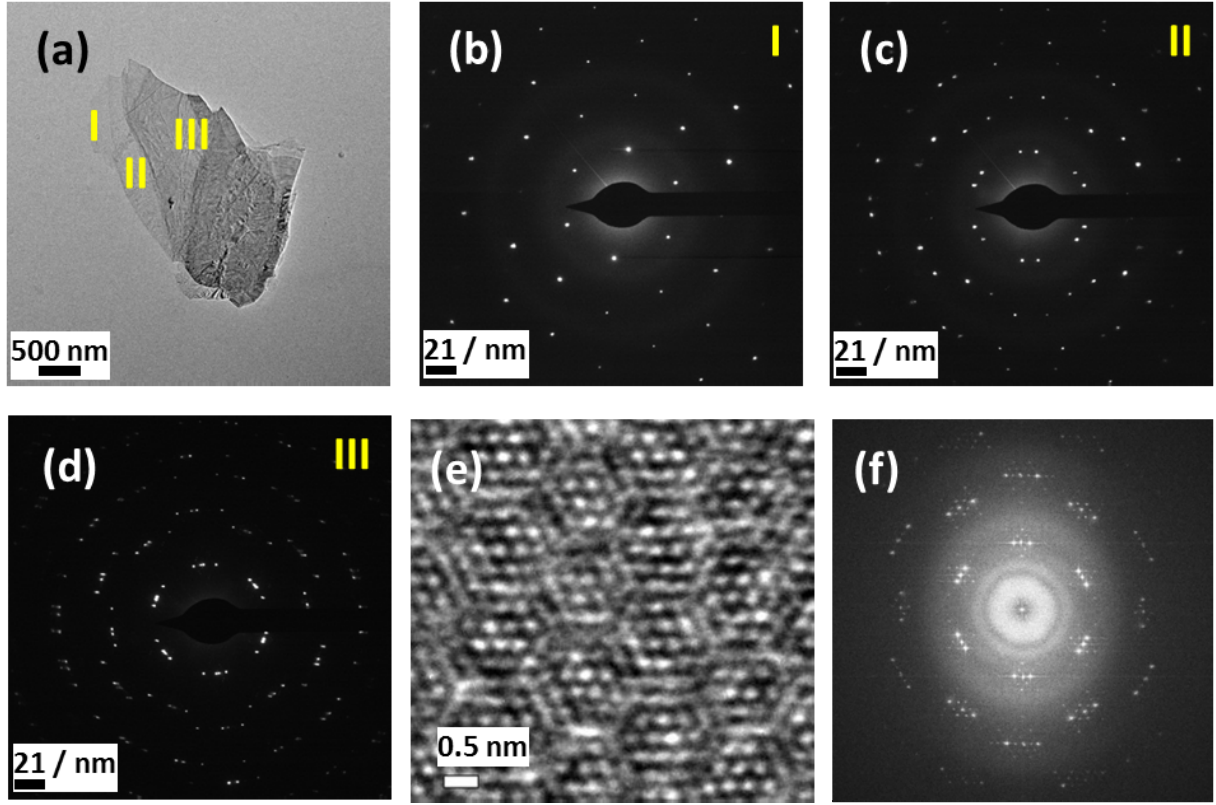

**FIG. 2.** (a) Transmission electron micrograph (TEM) of the s-MoS$_2$ multilayers. TEM electron diffraction patterns from (b) region I revealing the single crystal layer, (c) region II revealing the superposition of two single crystal layers with rotation, and (d) region III revealing the superposition of four single crystal layers with different rotations. The electron diffraction patterns reveal that the lattice parameter of s-MoS$_2$ is 0.32 nm. (e) An atomic-scale Moiré pattern of s-MoS$_2$ multilayers. (f) Fast Fourier transform (FFT) from the image of (e), indicating three different crystal layers with different rotational angles.

Figure 2(a) shows the morphology of the s-MoS$_2$ on a TEM grid where a single crystal layer region (I), a superposition of two single crystal layer with rotation (II), and superposition of four single crystal layers with rotation (III) are clearly resolved. The corresponding TEM electron diffraction patterns are shown in Fig. 2(b-d), respectively, showing the number of spots corresponding to the number of layers and the angular separation superimposed on the hexagonal lattice pattern. The electron diffraction patterns in Fig. 2(b-d) reveal that s-MoS$_2$ flakes retain the crystal symmetry and lattice constant, ~0.32 nm, when compared with that of x-MoS$_2$ in Fig. 3. A high-resolution TEM image is shown in Fig. 2(e) and its Fast Fourier transform (FFT) is shown in Fig. 2(f). The FFT shows that the Moiré pattern in the image is due to three different crystal layers superimposed with different angles.

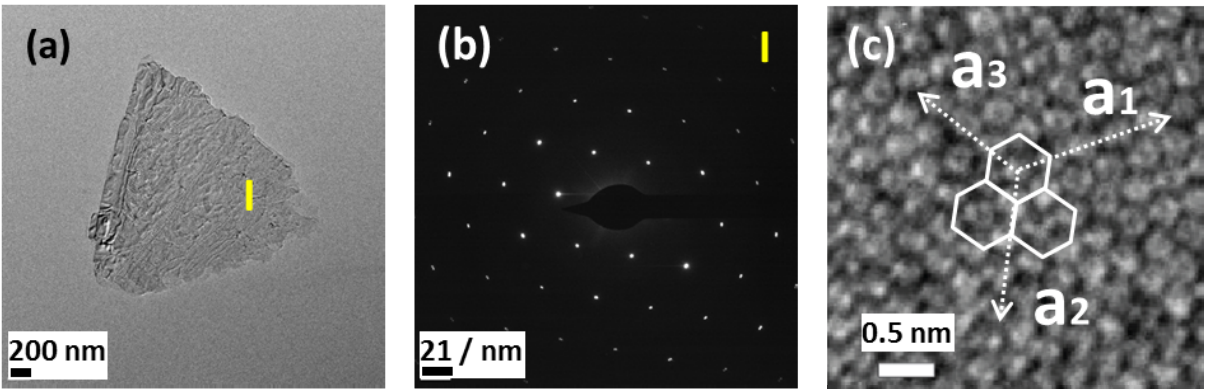

**FIG. 3**. (a) Transmission electron micrograph (TEM) of the x-MoS$_2$ multilayers. (b) Electron diffraction pattern from region I revealing the single crystal layer. (c) High-resolution (HR) TEM image of region I. A lattice constant of 0.32 nm can be obtained from (b) and (c).

For comparison, the morphology of x-MoS$_2$ on the TEM grid and the electron diffraction pattern with a high-resolution TEM image are shown in Fig. 3(a-c), respectively. The measurement confirms again a lattice parameter of x-MoS$_2$ of ~0.32 nm which is identical to that of s-MoS$_2$ and that of bulk MoS$_2$ [16]. The single-layer region is clearly resolved in the electron diffraction pattern, and the high-resolution atomic image clearly resolves the hexagonal crystal structure of single-layer MoS$_2$.

In summary, chemically synthesized MoS$_2$ transistors were fabricated and characterized for the first time, and compared with exfoliated MoS$_2$. The electronic and structural properties of chemically synthesized layers were found to be highly similar to exfoliated $MoS_2$. In particular, the transistor characteristics of s-$MoS_2$ were found to be almost identical to that of the x-$MoS_2$ in terms of FET mobility, subthreshold swing, on/off ratio, and contact resistance. TEM electron diffraction patterns and Raman measurements prove that the crystal symmetries and structural properties of s-$MoS_2$ are also identical to that of x-$MoS_2$. Though a number of issues need to be resolved before TMD crystals deliver high-performance transistors, this initial report shows that synthetic procedures can also realize high-quality channel material.

**ACKNOWLEDGEMENTS**

This work was supported in part by the Semiconductor Research Corporation (SRC), Nanoelectronics Research Initiative (NRI) and the National Institute of Standards and Technology (NIST) through the Midwest Institute for Nanoelectronics Discovery (MIND), the Office of Naval Research (ONR), and the National Science Foundation (NSF) award #1232191 monitored by Dr. A. Kaul, and the Air Force Office of Scientific Research (AFOSR) award #FA9550-12-1-0257 monitored by Dr. J. Hwang. We thank J. Jelenc for technical help in crystal growth, Slovenian Research Agency of the Republic of Slovenia for financial support, contract no. J1-2352 and the Centre of Excellence NAMASTE.